\documentclass{pasj00}
\draft

\begin{document}
\SetRunningHead{Y. Itoh et al.}{Molecular Hydrogen Emission 
from LkH$\alpha$ 264}
\Received{1999/12/31}
\Accepted{2000/01/01}

\title{Detection of Molecular Hydrogen Emission Associated with LkH$\alpha$ 264
\thanks{Based on data collected at Subaru Telescope, which is operated by
the National Astronomical Observatory of Japan}}

\author{Yoichi \textsc{Itoh}
\thanks{A guest investigator of the UK Astronomy Data Centre}}
\affil{Graduate School of Science and Technology, Kobe University,\\
1-1 Rokkodai, Nada, Kobe, Hyogo 657-8501}
\email{yitoh@kobe-u.ac.jp}
\author{Koji \textsc{Sugitani}}
\affil{Institute of Natural Sciences, Nagoya City University,\\
Mizuho-ku, Nagoya 467-8501}
\author{Katsuo \textsc{Ogura}}
\affil{Kokugakuin University, 4-10-28 Higashi, Shibuya-ku, 
Tokyo 150-8440}
\and
\author{Motohide \textsc{Tamura}}
\affil{National Astronomical Observatory of Japan,
2-21-1 Osawa, Mitaka, Tokyo 181-8588}


\maketitle

\begin{abstract}
We have detected emission of molecular hydrogen from a classical T Tauri
star, LkH$\alpha$ 264 in the $v=1-0~S(1)$ line at 2.122 $\micron$. 
The line velocity is coincident with the rest velocity of the star.
The line profile is well reproduced by a model in which
the line emanates from
material in a Keplerian rotating circumstellar disk.
Fluorescence by X-ray ionization and shock excitation due to 
accretion or a low-velocity wind are considered
for the emission mechanism of molecular hydrogen.
\end{abstract}

\section{Introduction}

A star is accompanied by circumstellar materials during its formation phase. 
Such materials are often recognized by their characteristic spectral features.
A circumstellar disk is detected by continuum excess in the infrared and
millimeter wavelengths.
Accretion from a circumstellar disk onto 
the stellar surface often produces continuum emission in the X-ray,
the ultraviolet wavelengths, and the optical wavelengths. It
also generates permitted emission lines, such as the H$\alpha$ line.
An outflow phenomenon is commonly traced by optical forbidden lines and 
by emission
lines of CO and other species in the radio wavelengths.
However, even though molecular hydrogen is considered to be the
most abundant element in
a young stellar disk system, it has so far been detected toward only few 
T Tauri stars (TTSs). This is because
H$_{2}$ does not have dipole moment.

Molecular hydrogen is excited in two ways.
One way is excitation by shock.
\citet{Herbst96} detected molecular hydrogen emission associated
with T Tau. Based primarily on the $2-1/1-0~S(1)$ ratio, 
they conclude that shock heating is a major excitation
mechanism for the emission.
The shock occurs where an outflow from the central binary
interacts with an ambient molecular cloud.
Otherwise, the shock may occur in an impact of accretion
gas onto a photosphere.

Another way of exciting molecular hydrogen
is fluorescence by ultraviolet or X-ray emission.
\citet{Thi} reported detection of H$_{2}$ emission lines toward GG Tau
in the mid-infrared wavelengths.
They claim that 
ultraviolet emission at the star-disk boundary excites
molecular hydrogen.
\citet{Weintraub} detected the H$_{2}~v=1-0~S(1)$ line toward 
a classical TTS (CTTS), TW Hya.
Because the flux of the line is in agreement with that predicted 
by an X-ray excitation model \citep{Maloney}, 
and because the line peak is coincident with the rest
velocity of the central star,
they attribute this emission to X-ray excitation.
Recently, \citet{Bary} and \citet{Bary03}
also detected a fluorescent H$_{2}~v=1-0 S(1)$ line from
three other TTSs (DoAr 21, GG Tau, and LkCa 15).

We present here high-resolution near-infrared spectroscopy of LkH$\alpha$ 264.
This star is a CTTS with
a spectral type of K5. Its apparent magnitude is $V \sim 12$
\citep{HerbigBell} with
a visual extinction of $A_{V}$ $\sim0.5$ mag.
The spectral energy distribution of LkH$\alpha$ 264 has 
a signature of a circumstellar disk in
the mid-infrared wavelengths \citep{Jaya01} and in the millimeter wavelengths
\citep{Itoh}.
The mass of the circumstellar disk is estimated to be 0.085 M$\solar$ \citep{Itoh}.
This star also exhibits a continuum excess in the blue region of the optical
wavelengths \citep{Valenti} and strong emission lines in 
the ultraviolet wavelengths (\cite{Gameiro93}; \cite{Costa}).
With these characteristics as well as time variations of the continuum and
emission lines (\cite{Lago}; \cite{Gameiro02}), 
LkH$\alpha$ 264 displays most of the known properties
of CTTSs.

LkH$\alpha$ 264 is associated with a high-latitude cloud,
MBM 12 \citep{Magnani}.
Sixteen TTSs are known so far to be associated with this cloud \citep{Ogura}.
The distance to the cloud has been thought to be around 65 pc
(\cite{Hobbs}; Hearty et al. 2000a).
However, recent studies of the stars
projected in the field of MBM 12 have indicated significantly 
larger distances around 300 pc (\cite{Luhman}; \cite{Andersson};
\cite{Straizys}).

This is one of our series of papers on the TTSs in MBM 12; we have
carried out detailed studies of those stars, in order to detect associated 
faint companions and disk structures, and to investigate their nature.
In this Letter, we focus on an emission line of
molecular hydrogen associated with LkH$\alpha$ 264. 
The other features detected by the
observations will be discussed in subsequent papers.

\section{Observations and Data Reduction}

High-resolution spectroscopic
observations were carried out on 2002 September 16 with 
the Infrared Camera and Spectrograph (IRCS)
on the Subaru Telescope at the summit of Mauna
Kea, Hawaii. IRCS has a 1024$\times$1024 InSb
array with a spatial scale of \timeform{0".060} pixel$^{-1}$ for echelle 
spectroscopy.
The echelle with the $K^{\ast}$ configuration
provides a wavelength coverage of 1.90 $\micron$ -- 2.45 $\micron$.
The width of the slit we used is \timeform{0".155}.
The resolving power is measured to be 20900 at 2.12 $\micron$,
using single and strong OH lines.
It corresponds to a velocity resolution of $\sim 14$ km s$^{-1}$.
The typical seeing size was \timeform{0".4} at 2 $\micron$
within a stable condition.
Four exposures were taken with the telescope dithered
approximately \timeform{3".2} along the slit for sky subtraction.
The total integration time is 600 seconds.
SAO 75672 (A0, $V = 9.1$) was observed for correcting
the effects of telluric absorption.
Exposures to a halogen lamp on and off were taken at the
end of the night.


The Image Reduction and Analysis Facility (IRAF) software was
used for all data reduction. 
First, a dithered pair of object frames were subtracted from each other,
then divided by a flat field. 
Next, we extracted an image of each order of the echelle spectra
using the APALL task with a "strip" option.
Then, each image was geometrically transformed to
correct the curvature of the slit image.
The solution of the wavelength calibration 
was derived from OH lines using the IDENTIFY task and the FITCOORDS task.
The wavelength can be varied as large as 0.3 $\rm \AA$ (4 km s$^{-1}$) by different
orders of the fitting function in the FITCOORDS task.
Individual spectra were extracted from the transformed images
using the APALL task.
The region where the
flux density of the object is more than 20\% of the peak flux density
at each wavelength was summed into a one-dimensional spectrum.
The object spectrum was divided by the standard star
spectrum,
and multiplied by a blackbody spectrum of a
temperature appropriate to the spectral type of the standard
star \citep{Tokunaga}.
The extracted spectra were then normalized and combined to produce a
final spectrum. 

\begin{figure}
  \begin{center}
    \FigureFile(85mm,85mm){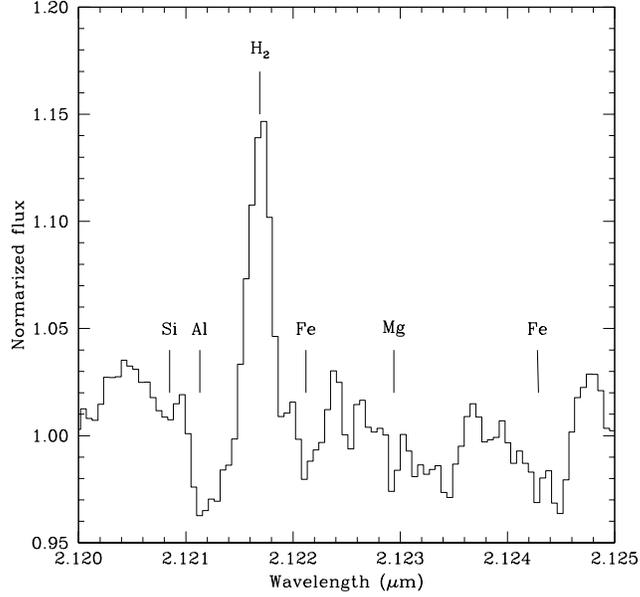}
  \end{center}
  \caption{Echelle spectrum of LkH$\alpha$ 264 around 
the H$_{2}~v=1-0~S(1)$ line. The spectrum is not corrected for $v_{LSR}$
of the star.}
\label{Lk264og}
\end{figure}

\section{Result}

A high-resolution spectrum of LkH$\alpha$ 264 around 2.122 $\micron$
is shown in Figure \ref{Lk264og}.
We detected the H$_{2}~v=1-0~S(1)$ line in emission and other metallic 
lines (Al, Fe, Mg, and Si) in absorption.

\subsection{Rest Velocity of the Star}

We measured $v_{LSR}$ of LkH$\alpha$ 264 with the Si line ($\lambda=$
2.1210 $\micron$) and the Fe line ($\lambda$=2.1244 $\micron$). These lines are
deep and sharp absorption lines near
the molecular hydrogen line \citep{Weintraub}.
$v_{LSR}$ is measured to be $-5.9$ km s$^{-1}$ with $\sigma=2.8$ km s$^{-1}$.
Change of $v_{LSR}$ due to the Earth's rotation during the observation 
was as small as 0.01 km s$^{-1}$.
Heliocentric velocity of LkH$\alpha$ 264 have been measured to be
+9.0 km s$^{-1}$ \citep{Herbig}, which corresponds to 
$v_{LSR}=+1.0\pm3.9$ km s$^{-1}$,
whereas Hearty et al. (2000b) derived $v_{LSR}=-4.2\pm2.5$ km s$^{-1}$.
$v_{LSR}$ derived from our observations is in agreement with that
of Hearty et al. (2000b).

\subsection{Molecular Hydrogen Emission Line}

The central wavelength of the molecular hydrogen emission line 
is measured to be
$2.121685\pm0.000008 \micron$ by a gaussian fitting.
It corresponds to $v_{LSR}=-5.1\pm1.2$ km s$^{-1}$.
Therefore, we conclude that the velocity of the molecular hydrogen line
is coincident with the rest velocity of the central star.
The emission line is symmetric and is not spatially resolved.
The FWHM of the line is measured to be $2.1\pm0.3\rm \AA$, which corresponds to
a width of $30\pm4$ km s$^{-1}$.
The equivalent width of the line is derived to be $-0.33\pm0.06\rm \AA$.
Compared with the $K$-band magnitude of the star, we estimate the flux of
the line to be $(3.7\pm0.6)\times10^{-14}$ erg s$^{-1}$ cm$^{-2}$,
or a luminosity of $(1.0\pm0.2)\times10^{-4}$ L$\solar$ for $d=300$ pc.
In the measurement above, we have estimated two kinds of uncertainties.
One is the uncertainties of the adjacent continuum level of the line.
The second is the deviations among the individual spectra.
We measured the line profile in the spectra before combining.
Finally, we considered larger values in these two values as the uncertainties
of the measurement.

We did not detect the H$_{2}~v=1-0~S(0)$ line and the H$_{2}~v=2-1~S(1)$ 
line, with 3$\sigma$ upper limits of $9.4\times10^{-15}$ erg s$^{-1}$ cm$^{-2}$ for
both lines. 
Therefore, the line ratios of $v=2-1~S(1)$ and of $v=1-0~S(0)$ to
$v=1-0~S(1)$ are less than 0.26 for LkH$\alpha$ 264. 

The mass of molecular hydrogen can be estimated from the flux of the line.
Assuming an LTE condition with an exciting temperature of 1500 K
and using equations (1) and (2) of \citet{Bary03},
we derive the mass of hot molecular hydrogen to be
$7\times10^{-7}$ M$\solar$.
However, this mass is thought to be only a small fraction of the total
disk mass.
Adopting a scale factor relating hot H$_{2}$ to the total disk mass
($10^{7} \sim 10^{9}$; \cite{Bary03}), we derive a total disk mass of
$7 \sim 700$ M$\solar$.
Such a heavy disk around a TTS is not realistic, because it should break
immediately by strong fragmentation. The small scale factor which indicates
efficient emission of molecular hydrogen should be applied
for LkH$\alpha$ 264.

An alternative explanation is that the distance to the object is 65 pc.
Using this distance with the same formulae above, we derive the mass
of the hot molecular hydrogen to be $3\times10^{-8}$ M$\solar$ and
the total disk mass to be $0.3 \sim 30$ M$\solar$.
Such estimates would be reasonable for a disk around LkH$\alpha$ 264,
especially at the lower value.

\subsection{Optical Spectra}

No optical forbidden emission lines have so far been detected from
LkH$\alpha$ 264 (e.g. \cite{Gameiro02}).
We investigated the optical spectra of the star in the ING Archive. 
The data were obtained on 2000 Oct. 12 using the Issac Newton Telescope.
The integration time is 600 sec.
The wavelength coverage is between 
3000 $\rm \AA$ and 9000 $\rm \AA$ with a spectral resolution of $\sim4000$.
We find no forbidden emission lines with an upper limit of $4.2\times10^{-15}$
erg s$^{-1}$ cm$^{-2}$ ( $1.2\times10^{-5}$ L$\solar$) for the [O I] line
at 6300 $\rm \AA$.
\citet{Hartigan} surveyed optical forbidden lines toward
TTSs in the Taurus molecular cloud. They detected the [O I] line for all CTTSs,
while they did not detect the line for all  weak-line TTSs 
with an upper limit of $\sim 4\times10^{-6}$ L$\solar$.
LkH$\alpha$ 264 does not show any forbidden lines as strong as those in
the CTTSs.
Therefore, this star does not have such an active outflow as
most CTTSs do.

\section{Discussion}

\subsection{Excitation Mechanism of the Molecular Hydrogen Emission Line}

Molecular Hydrogen can emit the $v=1-0~S(1)$ line through fluorescence
by X-ray or UV photons. The emission can also be due to shock caused
by accretion, by a low velocity wind, or by a high velocity jet. In
the following we will examine these possibilities.

\subsubsection{Fluorescent Emission}

Fluorescent emissions of molecular hydrogen excited by X-ray or UV
photons are found for several TTSs (\cite{Thi};
\cite{Weintraub}; \cite{Bary}; \cite{Bary03}).
\citet{Weintraub} detected the H$_{2}~v=1-0~S(1)$ emission line
toward TW Hya.
Since the velocity of the line is coincident with
that of the central star, they attribute the emission to 
fluorescence.
The same interpretation is made for DoAr 21 \citep{Bary}, GG Tau, and 
LkCa 15 \citep{Bary03}. 
The line velocity for LkH$\alpha$ 264
is also consistent with the velocity of the star.
Therefore, the line seems to be of fluorescent emission.
However, the flux ratio of the $v=2-1~S(1)$ line to the $v=1-0~S(1)$ line
is $\sim$ 0.5 for emission
induced purely by ultraviolet fluorescence \citep{Black}.
This ratio is measured to be
less than 0.26 for LkH$\alpha$ 264, indicating that the emission
is not induced by pure ultraviolet fluorescence.

X-ray excitation of molecular hydrogen has been investigated
for several kinds of objects.
\citet{Tine} consider an interstellar cloud heated by X-rays.
For some cases in the model, for instance $T=2000$ K and $n=10^{5}$ cm$^{-3}$,
the predicted line ratios of $v=2-1~S(1)$ and $v=1-0~S(0)$ to
$v=1-0~S(1)$ are consistent with the observed ratios.

If the H$_{2}$ emission line emanates from a circumstellar disk
by X-ray excitation,
we predict the $1-0~S(1)$ line intensity following \citet{Bary03}.
The X-ray luminosity of LkH$\alpha$ 264 
has been measured to be $10^{28.4}$ erg s$^{-1}$, assuming a distance to
the star of 65 pc (Hearty et al. 2000a). It corresponds to $10^{29.7}$ 
erg s$^{-1}$ for $d = 300$ pc.
Using the equations in \citet{Maloney},
we estimate the X-ray energy deposition rate per particle, $H_{X}$, to be
$8.5\times10^{-24}$ erg s$^{-1}$ at a distance of 10 AU from the source.
With Figure 6a of \citet{Maloney}, we find that the line intensity
would be $\sim 10^{-4.5}$ erg s$^{-1}$ cm$^{-22}$ str$^{-1}$ for
a plausible H$_{2}$ disk density of $n = 10^{5}$ cm$^{-3}$ and
a hydrogen column density between the source and the emitting gas
of $10^{22}$ cm$^{-2}$. 
The excitation temperature of H$_{2}$ is between 
1000 K and 2000 K \citep{Bary03}.
Assuming an annulus between 10 AU and 30 AU in a circumstellar disk
for the emitting region,
the line flux would be $\sim 2.1\times10^{-17}$ erg s$^{-1}$ cm$^{-2}$.
For $d=65$ pc, we find $H_{X} = 4.3\times10^{-25}$ erg s$^{-1}$, and
the line flux $\sim 4.5\times10^{-17}$ erg s$^{-1}$ cm$^{-2}$.
In either cases, the predicted line flux is three orders of magnitude smaller
than the observed values.

\subsubsection{Shock Excitation}

Another mechanism inducing molecular hydrogen emission is shock excitation.
Molecular hydrogen emission lines are often observed toward shock phenomena,
such as Herbig-Haro objects.
H$_{2}$ emission lines from a T Tau binary system are well interpreted by
shock excitation of outflow or accretion \citep{Herbst96}.
However, LkH$\alpha$ 264 does not have any signature of shock,
such as optical forbidden lines or associated Herbig-Haro objects 
or CO outflows.

First, we consider a shock by an unidentified
jet.
In this case, the jet should not extend over 60 AU, because the H$_{2}$
emission is not spatially resolved. Moreover, 
the jet should be highly inclined with respect to the line of sight, 
because the emission line is not largely blueshifted.
With such a jet, a circumstellar disk around the star would shade
the light of the central star.
Therefore, it is unlikely that a shock by an unidentified jet
makes the emission line.

Accretion of matter in a circumstellar disk
onto the central star may account for the line emission \citep{Herbst96}.
Such accretion generally redshifts a line.
Because the molecular hydrogen line of LkH$\alpha$ 264 does not
have a high velocity relative to the star, pole-on geometry is required
for this mechanism.

\citet{Hartigan} consider
shock excitation by wind associated with a circumstellar
disk for low-velocity components of optical forbidden lines
toward CTTSs. 
They revealed that such components often have small negative
radial velocities ($\sim 5$ km s$^{-1}$). Since the velocities
of the star and the emission line for LkH$\alpha$ 264
have relatively large uncertainties,
we cannot reject the possibility that the line is slightly blueshifted.

In summary, possible shock mechanisms inducing the molecular 
hydrogen emission are due to disk wind or accretion.
Further precise velocity measurements are required in order to determine
the excitation mechanism.

\subsection{The Line Profile of the Molecular Hydrogen Emission Line}

The line width of the molecular hydrogen line is significantly larger
than the resolution of the observations.
Even though we cannot determine whether molecular hydrogen is being
excited in fluorescent or by shock, we can infer that the emission is
associated with a circumstellar disk except for shock by accretion.
A line profile is broadened, if the emitting material
rotates in a circumstellar disk.
We construct a simple model for the line profile, 
following the model of \citet{Hartigan}.
We assume a Keplerian rotating disk.
A line profile generated by a ring of material in the disk has
a double-peak, with the two maxima occurring at $V_{max}=\pm V\sin i$,
where $V$ is the orbital velocity of the ring and $i$ is 
inclination.
We assume the surface brightness of the emitting material
in a power law according to the radius ($\propto r^{\nu}$).
The outer radius of the disk ($r_{out}$)
does not affect the line profile significantly.
We set an outer radius of 30 AU.
The line is smoothed by a gaussian function of the instrument profile
measured by OH lines.

\begin{figure}
  \begin{center}
    \FigureFile(160mm,160mm){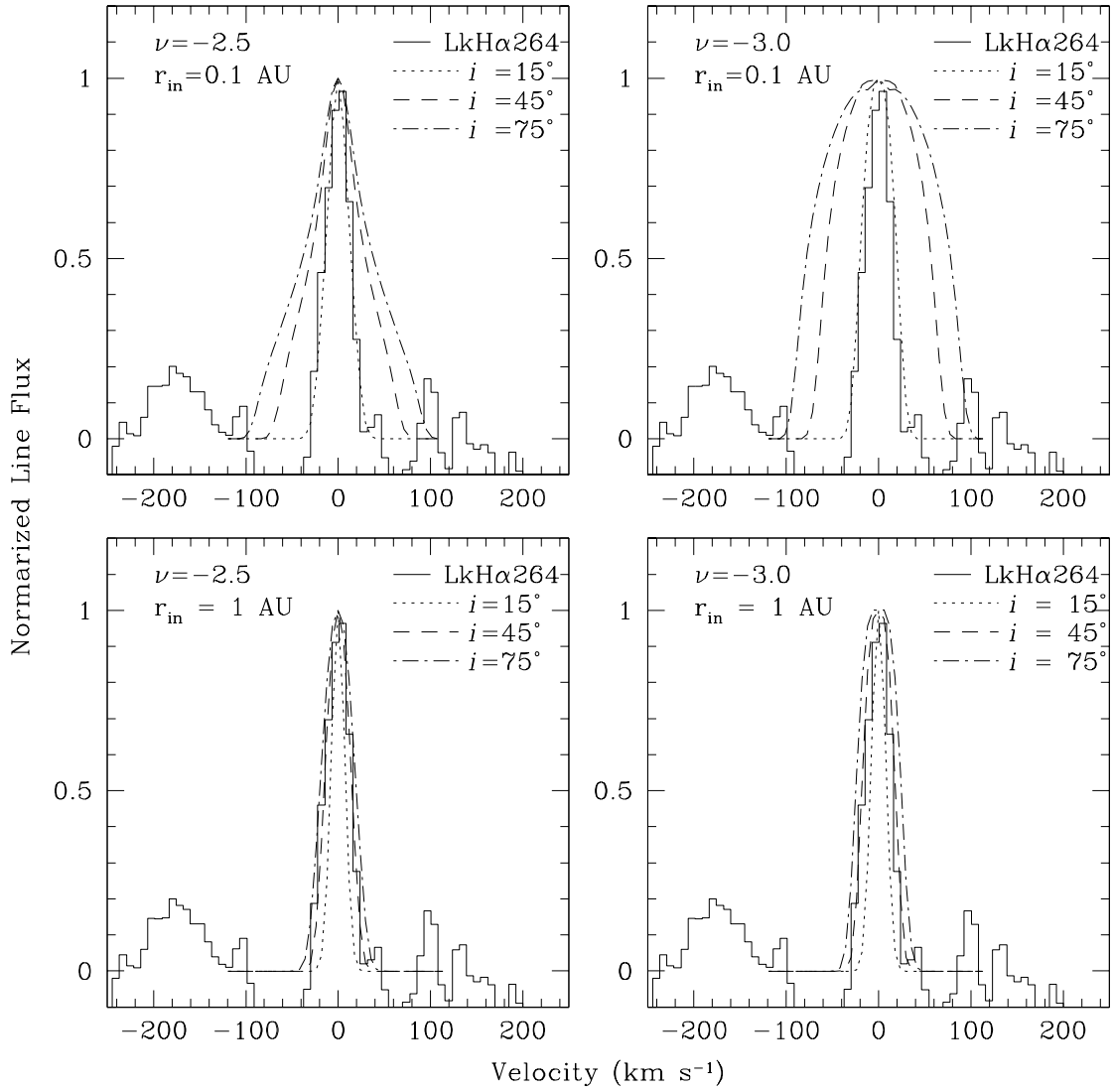}
  \end{center}
  \caption{The observed H$_{2}~v=1-0~S(1)$ emission line with the 
emission lines predicted by the models in which the emission emanates from
material in a circumstellar disk.}
\label{disk1}
\end{figure}

The predicted H$_{2}$ emission lines with the observed line are shown in 
Figure \ref{disk1}.
General behaviors of the line profiles are as follows; Pole-on geometry
makes a narrower profile than edge-on; A steep power law in brightness
generates a wider profile;
Emission from the region between 0.1 AU and 1 AU makes the line profile
wider or adds a wing to the profile.
The disk models that are consistent with the observed line profiles are
models with $\nu=-2.5$, $r_{in}$(the inner radius of the disk)$=0.1$ AU, 
$i=\timeform{15D}$,
with $\nu=-2.5$, $r_{in}=1$ AU, $i=\timeform{75D}$,
with $\nu=-3.0$, $r_{in}=0.1$ AU, $i=\timeform{15D}$, and
with $\nu=-3.0$, $r_{in}=1$ AU, $i=\timeform{45D}$.
\citet{Hartigan} found $\nu \sim -2.2$ for optical [O I] lines
for two CTTSs.
For LkH$\alpha$ 264, the models with $\nu=-2.2$ make narrower lines
than the observed line with any inner radius or any inclination.

An emission line from a circumstellar disk often has a double-peak.
The line tends to have a double-peak, if the inclination is large
and/or if the inner region of the disk emits a large portion of the 
emission. For example, the line is resolved in a double-peak for a
model with $\nu=-3.0$, $r_{in}=1$ AU, $r_{out}=10$ AU, and $i=\timeform{75D}$.
If the emission emanates not only from the inner region but also
from the outer region, a large negative number is required in $\nu$ for
a double-peak.
Since we assume a Keplerian rotating disk, $v\propto r^{1/2}$.
Therefore, a width of an annulus, $\Delta r$, in which 
the material rotates between $v$ and $v+\Delta v$ is $\propto r^{2}$.
An area of such an annulus is $2\pi r \Delta r \propto r^{3}$.
Therefore, in the case of $\nu > -3$, the outer region of the disk
emits a large portion of the emission, and the line tends to have a single-peak.
On the other hand, for $\nu < -3$, since the inner region emits a large portion
of the emission, the line tends to have a double-peak.
For example, the line is resolved in a double-peak for
a model with $\nu=-3.5$, $r_{in}=1$ AU, $r_{out}=30$ AU, and $i=\timeform{75D}$.
Though we cannot reject the possibility that the line has a double-peak with
a small velocity separation, two models above are, at least, inconsistent with
the observed line profile.

We considered that the X-ray induced H$_{2}$ emission may emanate
from an annulus between 10 AU and 30 AU in a disk. However,
the model with $r_{in}=10$ AU and $r_{out}=30$ AU produces
the line somewhat narrower than the observed line profile, with any
$i$ and $\nu$.
Nevertheless, because the temperature distribution of the disk, 
therefore $r_{in}$ and $r_{out}$ could vary with the disk shape,
we cannot reject the X-ray induced mechanism.

The spectral energy distribution of LkH$\alpha$ 264 indicates
the inner radius of the disk
to be 0.08 AU \citep{Itoh}.
We speculate, for the case of $r_{in}=1$ AU, 
that the circumstellar disk is thin and flat within 1 AU and
is flared beyond 1 AU. For such a disk, X-ray or shock by disk wind
much affects the disk surface beyond 1 AU, whereas little within 1 AU.
Otherwise, hydrogen exists within 1 AU in the form of an atom or be ionized.
The other explanation is that gas is depleted within 1 AU. 
Alternatively, for disk wind shock,
\citet{Safier} predicts that
forbidden lines and atomic hydrogen emission lines
emanate from a circumstellar disk within 1 AU, while molecular hydrogen
lines arise from a region beyond 1 AU.

~\\
~\\

We are grateful to H. Terada and R. Potter for help with the observations.
This research is partially based on data from the ING Archive.
Y. I. is supported by the Sumitomo Foundation.
This study is also supported by Grands-in-Aid from the Ministry of
Education, Culture, Sports, Science, and Technology of Japan
(14540228 for K. S. and Y. I., and 15540238 for K. O.).


\end{document}